\documentclass[reprint,superscriptaddress,showpacs,amsmath,amssymb,aps]{revtex4-1}

\usepackage{amsmath}
\usepackage{amssymb}
\usepackage{graphics}
\usepackage{graphicx}
\usepackage{dcolumn}
\usepackage{txfonts}
\usepackage[pdfpagemode=UseNone,pdfstartview=FitH,colorlinks=true,linkcolor=blue,urlcolor=blue,anchorcolor=blue,citecolor=blue]{hyperref}
\allowdisplaybreaks[4]

\begin{document}

\title{Magnon Kerr effect in a strongly coupled cavity-magnon system}
\author{Yi-Pu  Wang}
\thanks{These authors contributed equally.}
\affiliation{Quantum Physics and Quantum Information Division, Beijing Computational Science Research Center, Beijing, 100193 China}
\author{Guo-Qiang Zhang}
\thanks{These authors contributed equally.}
\affiliation{Quantum Physics and Quantum Information Division, Beijing Computational Science Research Center, Beijing, 100193 China}
\author{Dengke Zhang}
\affiliation{Quantum Physics and Quantum Information Division, Beijing Computational Science Research Center, Beijing, 100193 China}
\author{Xiao-Qing Luo}
\affiliation{Quantum Physics and Quantum Information Division, Beijing Computational Science Research Center, Beijing, 100193 China}
\author{Wei Xiong}
\affiliation{Quantum Physics and Quantum Information Division, Beijing Computational Science Research Center, Beijing, 100193 China}
\author{Shuai-Peng Wang}
\affiliation{Quantum Physics and Quantum Information Division, Beijing Computational Science Research Center, Beijing, 100193 China}
\author{Tie-Fu Li}
\thanks{litf@tsinghua.edu.cn}
\affiliation{Institute of Microelectronics, Tsinghua National Laboratory of
Information Science and Technology, Tsinghua University, Beijing 100084,
China}
\affiliation{Quantum Physics and Quantum Information Division, Beijing Computational Science Research Center, Beijing, 100193 China}
\author{C.-M. Hu}
\affiliation{Department of Physics and Astronomy, University of Manitoba, Winnipeg R3T 2N2, Canada}
\author{J. Q. You}
\thanks{jqyou@csrc.ac.cn}
\affiliation{Quantum Physics and Quantum Information Division, Beijing Computational Science Research Center, Beijing, 100193 China}

\date{\today}
\begin{abstract}
We experimentally demonstrate magnon Kerr effect in a cavity-magnon system, where magnons in a small yttrium iron garnet (YIG) sphere are strongly but dispersively coupled to the photons in a three-dimensional cavity. When the YIG sphere is pumped to generate considerable magnons, the Kerr effect yields a perceptible shift of the cavity's central frequency and more appreciable shifts of the magnon modes. We derive an analytical relation between the magnon frequency shift and the drive power for the uniformly magnetized YIG sphere and find that it agrees very well with the experimental results of the Kittel mode. Our study paves the way to explore nonlinear effects in the cavity-magnon system.
\end{abstract}


\maketitle

\section{Introduction}
Hybridizing two or more quantum systems can harness the distinct advantages of different systems to implement quantum information processors (see, e.g., Ref.\cite{Xiang,Kurizki}). Recently, a cavity-magnon system has attracted considerable attention~\cite{Huebl-13,Tabuchi-14,Zhang-14,Tobar-14,Hu-15,You-15,Hu-16}, because of the enhanced coupling between magnons in a yttrium iron garnet (YIG) single crystal and microwave photons in a high-finesse cavity. This hybrid system involves magnon polaritons~\cite{Cao-15,Hu-16-2}. Thus, a series of phenomena realized in other polariton systems~\cite{Carusotto-13,Kavokin}, including the Bose-Einstein condensation of exciton polaritons~\cite{Kasprzak-06,Deng-16} and the optical bistability in semiconductor microcavities~\cite{Baas-04}, can be explored using the magnon polaritons. Based on the strongly coupled cavity-magnon system, coherent interaction between a magnon and a superconducting qubit was realized~\cite{Tabuchi-15}, and magnon dark modes in a magnon gradient memory~\cite{Zhang-15} were utilized to store quantum information. When combined with spin pumping techniques, this cavity-magnon system provides a new platform to explore the physics of spintronics and to design useful functional devices~\cite{Hu-15,Hu-16}. Potentially acting as a quantum information transducer, microwave-to-optical frequency conversion between microwave photons generated by a superconducting circuit and optical photons of a whispering gallery mode supported by a YIG microsphere was also explored~\cite{Osada-16,Hisatomi-16,Haigh-15-2,Tang-arxiv}.
Furthermore, coherent phonon-magnon interactions relying on the effect of magnetostrictive deformation in a YIG sphere was demonstrated~\cite{Zhang-16}. Now, a versatile quantum information processing platform based on the coherent couplings among magnons, microwave photons, optical photons, phonons, and superconducting qubits is being established.

In this paper, we report an experimental demonstration of the magnon Kerr effect in a strongly coupled cavity-magnon system. The magnons in a small YIG sphere are strongly but dispersively coupled to the microwave photons in a three-dimensional (3D) cavity. When considerable magnons are generated by pumping the YIG sphere, the Kerr effect gives rise to a shift of the cavity's central frequency and yields more appreciable shifts of the magnon modes, including the Kittel mode~\cite{Kittel}, which holds homogeneous magnetization, and the magnetostatic (MS) modes~\cite{Walker-57,Walker-58,Bell}, which have inhomogeneous magnetization. We derive an analytical relation between the magnon frequency shift and the pumping power for a uniformly magnetized YIG sphere and find that it agrees very well with the experimental results of the Kittel mode. In contrast, the experimental results of MS modes deviate from this relation, which confirms the deviation of the MS modes from homogeneous magnetization.
To enhance the magnon Kerr effect, the pumping field is designed to directly drive the YIG sphere and its coupling to the magnons is strengthened using a loop antenna. Moreover, this pumping field is tuned very off-resonance with the cavity mode to avoid producing any appreciable effects on the cavity.
Our paper is a convincing study of a cavity-magnon system with magnon Kerr effect and paves the way to experimentally explore nonlinear effects in this tunable cavity-magnon system.

\begin{figure}[tbp]
  \centering
  \includegraphics[width=0.46\textwidth]{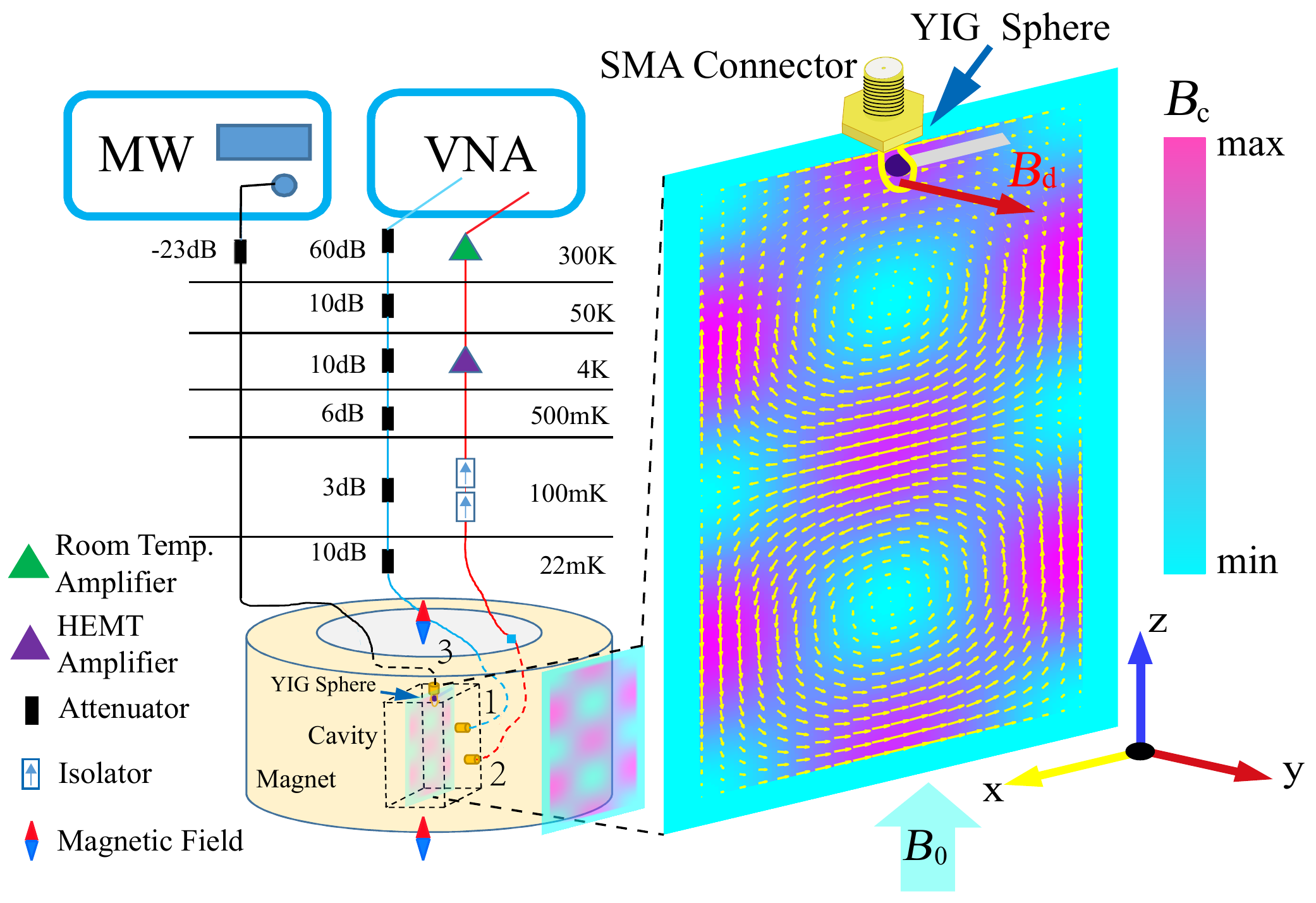}
  \caption{(color online). Schematic of the experiment setup. The 3D cavity is placed in the uniform magnetic field created by a superconducting magnet. Ports 1 and 2 are used for transmission spectroscopy and port 3 is for driving the YIG sphere via a superconducting microwave line with a loop antenna at its end. The total attenuation of the input port is 99dB. The right part shows the magnetic-field distribution of the cavity mode $\rm{TE}_{\rm{102}}$. The bias magnetic field, the drive magnetic field and the magnetic field of the $\rm{TE}_{\rm{102}}$ mode are mutually perpendicular at the site of the YIG sphere, where the magnetic field of the $\rm{TE}_{\rm{102}}$ mode is maximal. }
  \label{fig:1}
\end{figure}

\section{Experimental setup}
The experimental setup is diagrammatically shown in Fig.~\ref{fig:1}. The 3D cavity is made of oxygen-free copper with inner dimensions of $44.0\times20.0\times6.0~\rm{mm}^{3}$ and contains three ports labeled as 1, 2 and 3 (here ports 1 and 2 are used for transmission spectroscopy and port 3 is for loading the drive field). The frequency of the cavity mode $\rm{TE}_{\rm{102}}$ that we use is  $\omega_{\rm{c}}/2\pi=10.1$~GHz.  A samll YIG sphere of diameter $1$~mm is glued on an inner wall of the cavity at the magnetic-field antinode of the $\rm{TE}_{\rm{102}}$ mode (c.f., the magnetic-field intensity distribution of this mode marked by colored shades in Fig.~\ref{fig:1}). We apply a static magnetic field generated by a superconducting magnet to magnetize the YIG sphere. This bias magnetic field is tunable in the range of 0 to 1~T, so the given frequency of the Kittel mode (i.e., the ferromagnetic resonance mode) ranges from several hundreds of megahertz to $28$~GHz. The cavity is placed in a BlueFors LD-400 dilution refrigerator under a cryogenic temperature of $22$~mK. The spectroscopic measurement is carried out with a vector network analyzer by probing the transmission of the cavity. A drive tone supplied by a microwave source can directly drive the YIG sphere via a superconducting microwave line going through port 3. Moreover, a loop antenna is attached to the end of the superconducting microwave line near the YIG sphere (see Appendix~\ref{AppA}), so as to strengthen the coupling between the drive field and the YIG sphere. Here the driving magnetic field $B_{\rm{d}}$, the bias magnetic field $B_0$ (which is aligned along the hard magnetization axis $[100]$ of the YIG sphere), and the magnetic field $B_{\rm{c}}$ of the $\rm{TE}_{\rm{102}}$ mode are orthogonal to each other at the site of the YIG sphere. Also, a series of attenuators and isolators is used to prevent thermal noise from reaching the sample and the output signal is amplified by two low-noise amplifiers at the stages of 4 K and room temperature, respectively.
\section{Strong coupling regime}
\begin{figure}
\centering
 \includegraphics[width=0.48\textwidth]{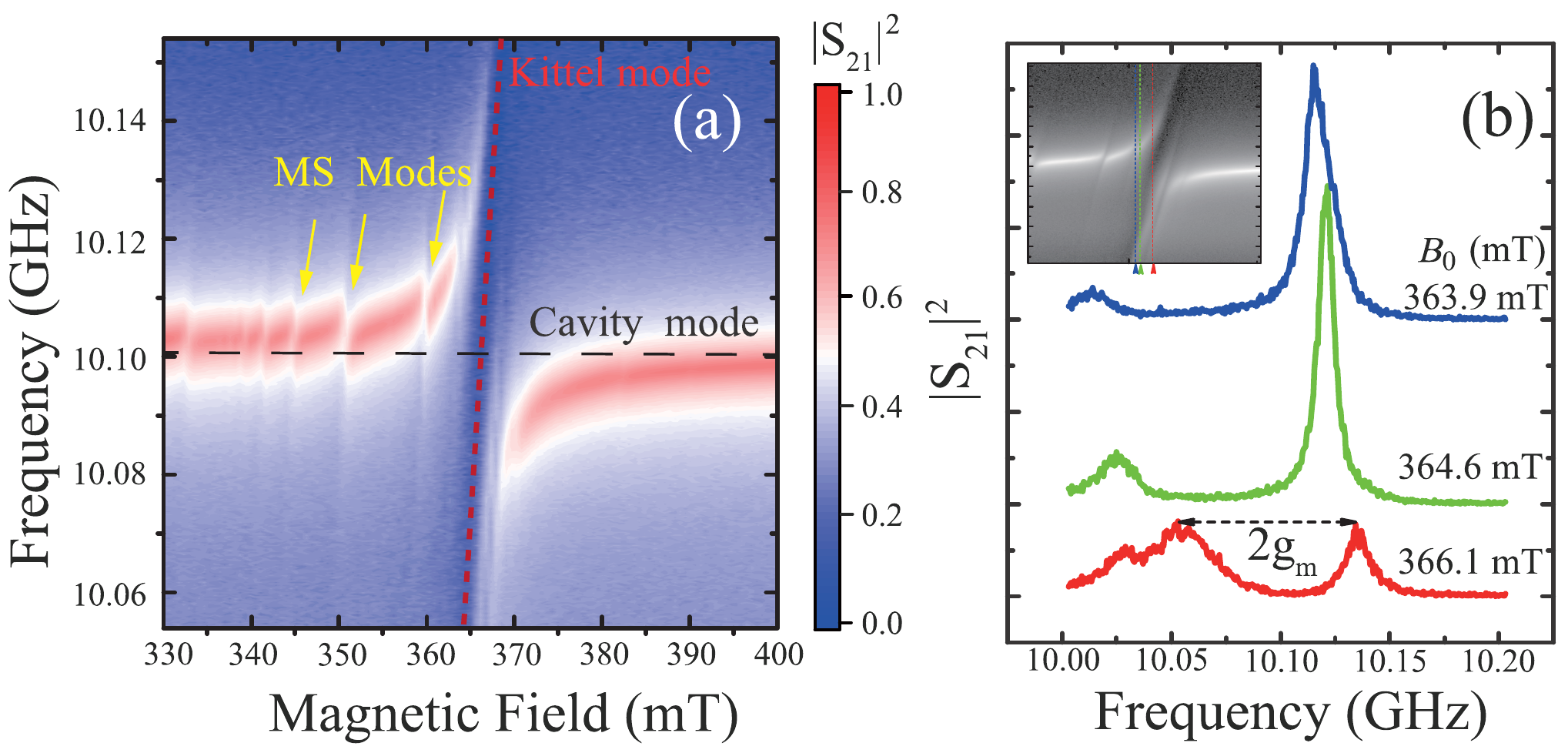}
 \caption{(color online). (a) Transmission spectrum for the normal-mode splitting measured as a function of the bias magnetic field and the probe microwave frequency. The large anticrossing indicates strong coupling between the Kittel mode and the cavity mode $\rm{TE}_{\rm{102}}$. The small splittings are due to the MS modes coupled with the cavity mode. (b) Transmission spectrum at three values of the bias magnetic field. The curves are offset vertically for clarity.}
 \label{fig:2}
\end{figure}

We first measure the transmission spectrum of the cavity containing the YIG sphere, without applying a drive field on the YIG sphere. The transmission spectrum as a function of  the probe microwave frequency and $B_{\rm{0}}$ are recorded by vector network analyzer [see Fig.~\ref{fig:2}(a)].
At the point where the Kittel mode is resonant with the cavity mode $\rm{TE}_{\rm{102}}$, a distinct
anti-crossing of the two modes occurs, indicating strong coupling between them. Some other small splittings are due to the couplings between the cavity mode and the MS modes in the YIG sphere.
The coupling strength between the Kittel mode and the cavity mode $\rm{TE}_{\rm{102}}$ is found to be $g_{\rm{m}}/2\pi=42\rm~{MHz}$ from the magnon polariton splitting at the resonance point [see the red cure in Fig.~\ref{fig:2}(b)]. By fitting the measured transmission spectrum~\cite{You-15}, the cavity-mode linewidth~$\kappa/2\pi\equiv(\kappa_{\rm{1}}+\kappa_{\rm{2}}+\kappa_{\rm{int}})/2\pi$ and the Kittel-mode linewidth $\gamma_{\rm{m}}/2\pi$ are determined to be $2.87$~MHz and $24.3$~MHz, respectively. Here $\kappa_1$ ($\kappa_2$) is the loss rate due to the port 1 (2) and $\kappa_{\rm{int}}$ is due to the intrinsic loss of the cavity. The obvious increase in the Kittel mode damping rate compared with the previous work~\cite{Tabuchi-14,Zhang-14} is due to the antenna close to the YIG sphere, which acts as an additional decay channel.
Note that all the linewidths throughout the paper are defined as the full width at half maximum (FWHM). Because $g_{\rm{m}}>\kappa,\gamma_{\rm{m}}$, the hybrid system falls in the strong-coupling regime with a cooperativity $C \equiv 4g_{\rm{m}} ^{2}/\kappa\gamma_{\rm{m}}=101.$

\section{Results and analysis of the dispersive measurement}
\subsection{Dispersive measurement}
\begin{figure}
\centering
 \includegraphics[width=0.45\textwidth]{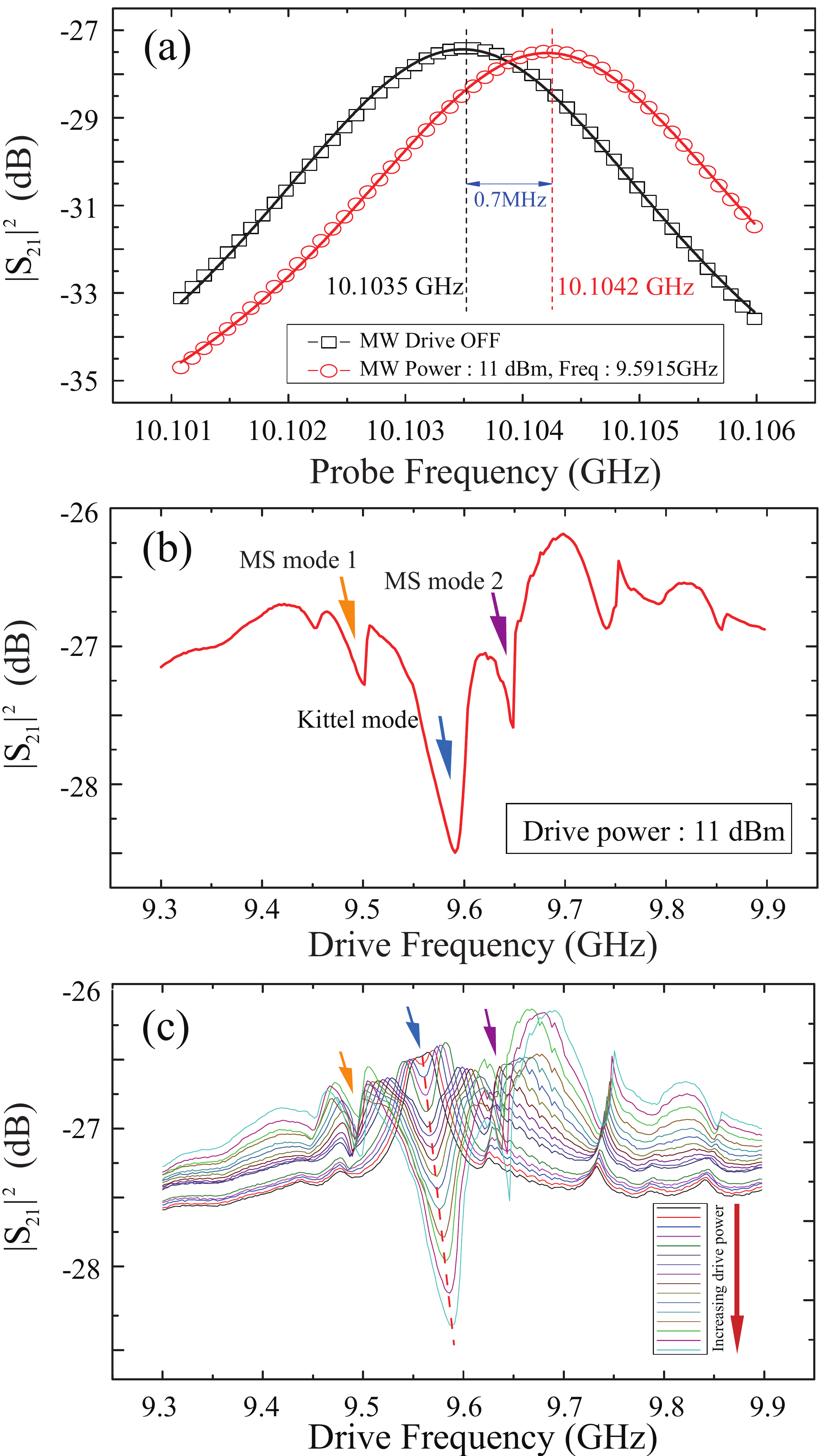}
 \caption{(color online). (a)~Central frequency shift of the cavity mode $\rm{TE}_{\rm{102}}$ when the drive field is on (red curve) and off (black curve), respectively. (b)~Transmission spectrum of the cavity measured as a function of the drive-field frequency. The blue arrow indicates the response of the Kittel mode, whereas the orange and purple arrows indicate the MS modes 1 and 2, respectively. (c)~Transmission spectrum of the cavity measured as a function of the drive frequency by successively increasing the driving power. The probe field is fixed at $10.1035$~GHz in both (b) and (c).}
 \label{fig:3}
\end{figure}

 We tune the static bias magnetic field $B_0$ to $346.8$~mT, yielding about 9.55 GHz for the frequency of the Kittel mode. As shown in Fig.~\ref{fig:3}(a), we first measure the transmission spectrum of the cavity (i.e., the black curve) by tuning the frequency of the probe field, but without the drive field on the YIG sphere. The measured central frequency of the cavity mode is 10.1035 GHz, which has a frequency shift of about 3 MHz compared with the intrinsic frequency of 10.1003 GHz of the $\rm{TE}_{\rm{102}}$ mode of an empty cavity. This cavity mode has a detuning $\Delta/2\pi\approx 550$~MHz from the Kittel mode. Because $\Delta>10g_{\rm{m}}$, the coupled hybrid system is in the dispersive regime.  We then measure the transmission spectrum of the cavity by both tuning the frequency of the probe field and applying a drive field on the YIG sphere in resonance with the Kittel mode. The measured red curve corresponds to the drive power of 11~dBm. This transmission spectrum has a central frequency of $10.1042$~GHz, with a frequency shift of about $0.7$~MHz from the measured central frequency without applying a drive field.

Figures~\ref{fig:3}(b) and \ref{fig:3}(c) show the measured transmission spectra by tuning the frequency of the drive field, where the frequency of the probe field is fixed at the central frequency of $10.1035$~GHz of the cavity containing the YIG sphere. The probe field power is -129 dBm. The corresponding average cavity probe photon number can be estimated by~\cite{Marquardt}
\begin{equation}\label{eq1}
\bar{n}=\frac{\kappa_1 P_{\rm{p}}}{\hbar\omega_{\rm{p}}[\Delta_{\rm{p}}^2+(\kappa/2)^2]},
\end{equation}
where $P_{\rm{p}}$ is the probe field power and $\Delta_{\rm{p}}=\omega_{\rm{p}}-\omega_{\rm{c}}$. In our experiment, it is measured that $\kappa_1/2\pi=0.70$~MHz, i.e., $\kappa_1\sim \kappa/4$. Also, the probe field frequency $\omega_{\rm{p}}$ is tuned in resonance with the cavity mode $\rm{TE}_{\rm{102}}$. Then, the average cavity probe photon number is reduced to $\bar{n}=P_{\rm{p}}/(\hbar\omega_{\rm{p}}\kappa)\approx 1$.
Here the probe tone is chosen extremely weak, so as to avoid producing any appreciable effects on the system. In Fig.~\ref{fig:3}(b), the power of the drive field is 11~dBm. It can be seen that
when the frequency of the drive microwave field is resonant with the Kittel mode, the transmission coefficient has a large decrease at $9.59$~GHz (see the main dip indicated by a blue arrow), caused by the shift of the central frequency of the cavity mode. The dips indicated by orange and purple arrows correspond to two different MS modes.
In addition, we vary the power of the drive field from -5 to 10 dBm in Fig.~\ref{fig:3}(c) and observe two interesting features, i.e., when increasing the drive power, the main dip becomes deeper successively and it simultaneously shifts rightwards. This reveals that the Kittel mode has a blue shift with the increase in the drive power. The responses of MS modes are similar.

\subsection{Origin of the Kerr term}
For a YIG sphere uniformly magnetized by an external magnetic field along the $z$ direction, when the magnetization is saturated, the induced internal magnetic field includes the demagnetizing field~\cite{Blundell}~$\mathbf{H}_{\rm{de}}=-\textbf{M}/3$ and the anisotropic field~\cite{Macdonald,Stancil} $\mathbf{H}_{\rm{an}}=-(2K_{\rm{an}}/M^{2})M_{\rm{z}}$, where $\textbf{M}\equiv(M_{x},M_{y},M_{z})$ is the magnetization, $M$ is the saturation magnetization and $K_{\rm{an}}$ is the first-order magnetocrystalline anisotropy constant of the YIG sphere. When both the Zeeman energy and the magnetocrystalline anisotropic energy are included (see Appendix~\ref{AppB}), the Hamiltonian of the YIG sphere in the magnetic field $B_{0}$ is given by (setting $\hbar=1$)
\begin{eqnarray}\label{eq2}
H_{\rm{m}}=-\gamma B_{0}S_{z}-\frac{\mu_{0}\gamma^{2}K_{\rm{an}}}{M^{2}V_{\rm{m}}}S_{z}^{2},
\end{eqnarray}
where $\gamma/2\pi=28$~GHz/T is the gyromagnetic ratio, $\mu_{0}$ is the vacuum permeability and $S_{z}=M_{z}V_{\rm{m}}/\gamma$ is a macrospin operator of the YIG sphere, with $V_{\rm{m}}$ being the volume of the YIG sample.
The macrospin operator $S_{z}$ is related to the magnon operators via the Holstein-Primakoff transformation~\cite{Holstein}: $S_{z}=S-b^{\dag}b$, where ${b}^{\dag}$($b$) is the magnon creation (annihilation) operator.

When including the drive field, the cavity mode, and the interaction between the cavity photon and the magnon, the total Hamiltonian of the coupled hybrid system is (see Appendix~\ref{AppB})
\begin{eqnarray}
H&=&\omega _{\rm{c}} a^\dag  a + \omega _{\rm{m}} b^\dag  b + K b^{\dag}  bb^\dag  b \nonumber\\
&&+ g_{\rm{m}} (a^\dag  b + ab^\dag)
+\Omega _{\rm{d}} (b^\dag  e^{ - i\omega _{\rm{d}} t}  + be^{i\omega _{\rm{d}} t} ),
\label{hamiltonian}
\end{eqnarray}
where $\hat{a}^{\dag}$ $(\hat{a})$ is the creation (annihilation) operator of the cavity photons at frequency $\omega_{\rm{c}}$,
$Kb^{\dag}  bb^{\dag}  b$ represents the Kerr effect of magnons owing to the magnetocrystalline anisotropy in the YIG sphere, with $K=\mu_{0}K_{\rm{an}}\gamma^{2}/(M^{2}V_{\rm{m}})$,
$\Omega_{\rm{d}}$ (i.e., the Rabi frequency) denotes the strength of the drive field, and $\omega_{\rm{d}}$ is the drive field frequency. Thus, our experimental setup provides a strongly coupled cavity-magnon system with the magnon Kerr effect, which is an extension of the cavity-magnon system without the nonlinear effect~\cite{Soykal}. Note that $K$ is inversely proportional to $V_m$, so the Kerr effect can become important when using a small YIG sphere.

\subsection{Cavity and magnon frequency shifts}
Below we study the case of considerable magnons generated by the drive field. Because the coupled hybrid system is in the dispersive regime, its effective Hamiltonian can be written as (see Appendix~\ref{AppC})
\begin{eqnarray}\label{s}
H_{\rm{eff}}&=&\left[\omega_{\rm{c}}+\frac{g_{\rm{m}}^2}{\Delta}+\frac{2g_{\rm{m}}^2}{\Delta^2}K\langle b^{\dag}b\rangle\right]a^{\dag}a \nonumber\\
&&+\left[\omega_{\rm{m}}-\frac{g_{\rm{m}}^2}{\Delta}+\left(1-\frac{2g_{\rm{m}}^2}{\Delta^2}\right)K\langle b^{\dag}b\rangle\right]b^{\dag}b   \nonumber\\
&&+\Omega'_{\rm{d}}(b^{\dag}e^{-i\omega_{\rm{d}}t}+be^{i\omega_{\rm{d}}t}),
]
\end{eqnarray}
with the effective Rabi frequency $\Omega'_{\rm{d}}$ given by
\begin{equation}
\Omega'_{\rm{d}}=\bigg[1-\frac{1}{2(\omega_{\rm{c}}-\omega_{\rm{d}})}
\bigg(\frac{g_{\rm{m}}^{2}}{\Delta}+\frac{2g_{\rm{m}}^{2}}{\Delta^{2}}K
\langle b^{\dag}b\rangle\bigg)\bigg]\Omega_{d},
\end{equation}
where $\Delta=\omega_{\rm{c}}-\omega_{\rm{m}}$. Due to the coupling between the cavity and the YIG sphere, the cavity frequency shifts from the intrinsic cavity mode frequency $\omega_{\rm{c}}$ to $\omega_{\rm{c}}+g^{2}_{\rm{m}}/\Delta$, with $g^{2}_{\rm{m}}/\Delta$ being the dispersive shift. The measured central frequency of 10.1035~GHz corresponds to $\omega_{\rm{c}}+g^{2}_{\rm{m}}/\Delta$.
When pumping the YIG sphere with a drive field, the magnon number $\langle b^{\dag}b\rangle$ increases. Then the cavity frequency has an additional blue shift of $\Delta_{\rm{c}}=(2g_{\rm{m}}^2/\Delta^2)K\langle b^{\dag}b\rangle$ due to the Kerr effect. Also, the Kerr effect yields a blue shift to the magnon frequency, $\Delta_{\rm{m}}=(1-2g_{\rm{m}}^2/\Delta^2)K\langle b^{\dag}b\rangle\approx K\langle b^{\dag}b\rangle$. Both cavity frequency shift and magnon frequency shift due to the Kerr effect have a similar trend depending on $\langle b^{\dag}b\rangle$, which is related to the drive power.

\begin{figure}
  \includegraphics[width=0.46\textwidth]{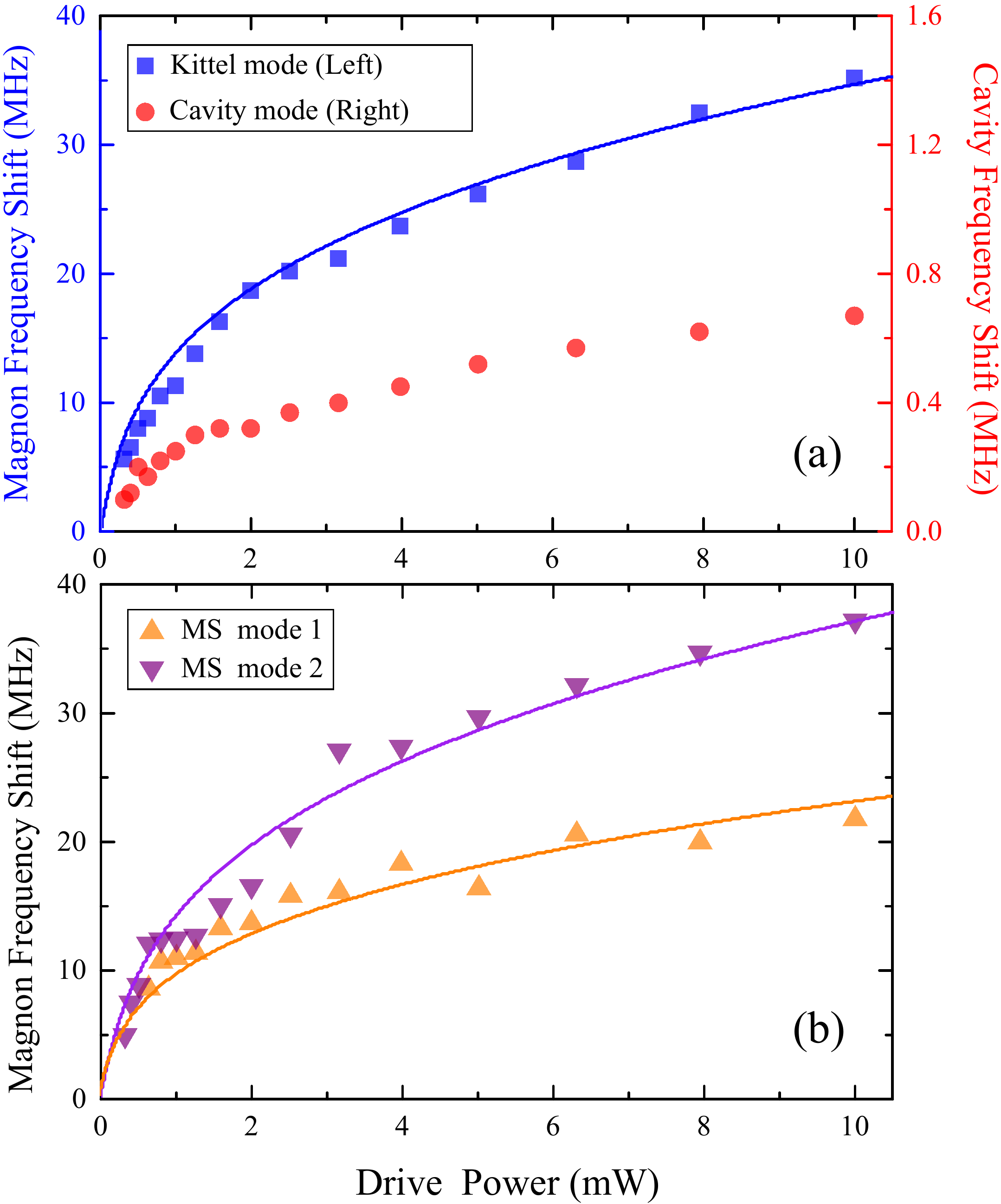}
  \caption{(color online). (a)~Frequency shift of the Kittel mode (blue square) and the central frequency shift of the cavity mode $\rm{TE}_{\rm{102}}$ (red circle) measured at various values of the drive power. The blue fitting curve for the Kittel mode is obtained using Eq.~(\ref{eq4}). (b)~Frequency shifts of MS mode 1 (orange up-triangle) and MS mode 2 (purple down-triangle) measured at various values of the drive power. The corresponding orange and purple fitting curves also are obtained using Eq.~(\ref{eq4}). The frequency shifts of the Kittel mode, MS mode 1 and MS mode 2 are here referenced from 9.5526, 9.4758, 9.6174GHz, respectively.}
  \label{fig:4}
\end{figure}

\section{Relation between the magnon frequency shift and the drive power}
In Fig.~\ref{fig:4}, we extract the Kerr-effect-induced frequency shifts of the magnon as well as the central frequency shift of the cavity mode at each given drive power $P$. From Fig.~\ref{fig:4}(a), it is clear that both the Kittel-mode frequency shift and the cavity central frequency shift indeed have similar behaviors depending on the drive power, as predicted above.
We also see that all the frequency shifts exhibit nonlinear dependence on the drive power.
As given in Appendix~\ref{AppD}, we derive an analytical relation between the magnon frequency shift $\Delta_{\rm{m}}$ and the drive power $P$ using a Langevin equation approach,
\begin{equation}
\bigg[\Delta_{\rm{m}}^2+\bigg(\frac{\gamma_{\rm{m}}}{2}\bigg)^2\bigg]\Delta_{\rm{m}}-cP=0,
\label{eq4}
\end{equation}
where $c$ is a characteristic parameter reflecting the coupling strength of the drive field with the magnon mode. For the Kittel mode, we have already measured its linewidth $\gamma_{\rm{m}}/2\pi=24.3~\rm MHz$. We use Eq.~(\ref{eq4}) to fit the experimental results of the Kittle mode. As shown in Fig.~\ref{fig:4}(a), the obtained theoretical (blue) curve fits very well with the experimental data, where $c=(2\pi)^{3}\times4.7\times10^{24}~\rm{kg^{-1} m^{-2} }$.

For the MS modes, we have two unknown parameters, the MS mode linewidth $\gamma_{\rm{m}}$ and the parameter $c$. We manage to fit the experimental data in Fig.~\ref{fig:4}(b) with $\gamma_{\rm{m}}=15~\rm MHz$ and $c=(2\pi)^{3}\times1.35\times10^{24}~\rm{kg^{-1} m^{-2} }$ for MS mode 1 (orange curve), and with $\gamma_{\rm{m}}=30~\rm MHz$ and $c=(2\pi)^{3}\times6\times10^{24}~\rm{kg^{-1} m^{-2} }$ for MS mode 2 (purple curve). Note that the theoretical curves do not fit the experimental data of the MS modes so well as those of the Kittel mode, especially in the region around the threshold power [see the region of 1-3~mW in Fig.~\ref{fig:4}(b)]. In fact, as a collective mode of spins with a zero wavevector, the Kittel mode is the uniform precession mode with homogeneous magnetization, whereas
the MS modes are nonuniform precession modes holding inhomogeneous magnetization and have a spatial variation comparable to the sample dimensions~\cite{Walker-58,Bell,Stancil}. The appreciable deviations of the experimental data from the theoretical fitting curves are due to the inhomogeneous magnetization of the MS modes.

Note that when the drive power is small, $\gamma_{\rm{m}}\gg\Delta_{\rm{m}}$, so Eq.~(\ref{eq4}) reduces to
\begin{equation}
\left(\frac{\gamma_{\rm{m}}}{2}\right)^2\Delta_{\rm{m}}-cP=0,
\end{equation}
i.e., the magnon frequency shift depends linearly on the drive power in the small drive power limit. When the drive power becomes sufficiently large, $\Delta_{\rm{m}}\gg\gamma_{\rm{m}}$, and then Eq.~(\ref{eq4}) reduces to
\begin{equation}
\Delta_{\rm{m}}^3-cP=0.
\end{equation}
It yields
\begin{equation}
\Delta_{\rm{m}}=(cP)^{1/3},
\end{equation}
i.e., in the large drive power limit, the magnon frequency shift depends linearly on the cubic root of the drive power. These limit results are consistent with the previous work in Ref.~\cite{Hu-09}, where there is a threshold power separating the small and large driving power regions.

A dispersive measurement on the cavity transmission was implemented at room temperature in Ref.~\cite{Haigh-15}, but the cavity's central frequency shift due to the magnon Kerr effect was not observed. In Ref.~\cite{Haigh-15}, the drive field was applied on the cavity rather than the YIG. This is different from our setup in which the YIG sphere is directly pumped by the drive field and the nonlinear effect of large-amplitude spin waves can be induced~\cite{Cash-96}. Moreover, choosing a suitable angle between the external magnetic field and the crystalline axis is also important to observe the magnon Kerr effect because the value of the Kerr coefficient $K$ strongly depends on this angle~\cite{Melkov}. In our case, the bias magnetic field $B_0$ is aligned along the hard magnetization axis [100] of the YIG sphere, which gives rise to the largest $K$. Furthermore, our experiment is implemented at a cryogenic temperature where the magnetocrystalline anisotropy constant $K_{\rm an}$ (so the Kerr coefficient $K$) is several times larger than that at room temperature~\cite{Huebl-13,Stancil}. These may be the reasons why appreciable Kerr effect was not observed in Ref.~\cite{Haigh-15}.

\section{conclusion}
We have realized a strongly coupled cavity-magnon system with magnon Kerr effect. By directly pumping the YIG sphere
with a drive field, we have demonstrated the Kerr-effect-induced central frequency shift of the cavity mode as well as the frequency shifts of the Kittel mode and MS modes.
An analytical relation between the magnon frequency shift and the pumping power for a uniformly magnetized YIG sphere is derived, which agrees very well with the experimental results of the Kittel mode. In contrast, the experimental results of MS modes deviate from this relation owing to the spatial variations of the MS modes over the sample. We can use this relation to characterize the degrees of deviation of the MS modes from the homogeneous magnetization.
Our setup can provide a flexible and tunable platform to further explore nonlinear effects of magnons in the cavity-magnon system. Moreover, this coupled hybrid system involves magnon polaritons. It can be used to explore a series  of phenomena realized in other polariton systems~\cite{Carusotto-13,Kavokin}.

\begin{acknowledgments}
This work was supported by the National Key Research and Development Program of China (Grant No.~2016YFA0301200), the MOST 973 Program of China (Grant No.~2014CB848700), and the NSAF (Grant No.~U1330201 and No. U1530401).
C.M.H. was supported by the NSFC (Grant No.~11429401).
\end{acknowledgments}

\appendix

\section{CAVITY USED IN THE EXPERIMENT}\label{AppA}

\begin{figure}[htbp]
 \centering
\includegraphics[width=0.4\textwidth]{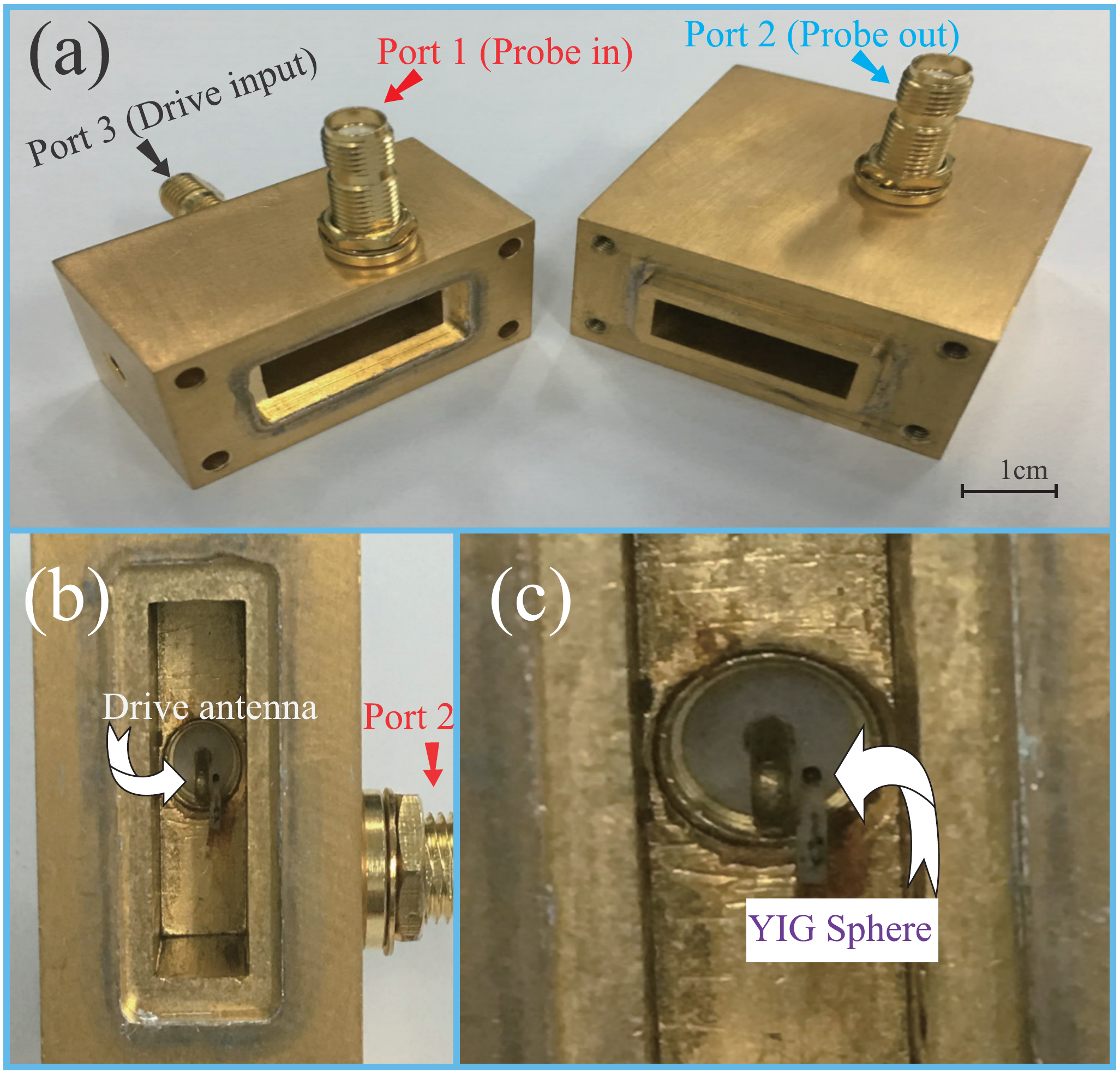}
\caption{(a) The 3D cavity, which is made of oxygen-free copper and plated with gold.  (b) The drive antenna, which is connected to a superconducting microwave line that goes into the cavity via port 3. (c) The small YIG sphere, which has a diameter of 1~mm and is placed near the antenna and glued on the inner wall of the cavity.}
\label{fig:5}
\end{figure}

Figure.~\ref{fig:5} shows the three-dimensional (3D) rectangular cavity used in our experiment. It has inner dimensions of $44.0\times20.0\times6.0~\rm{mm}^{3}$ and contains three ports. The Port 1 (2) is used for the probe field into (out of) the cavity and the Port 3 is used for inputting the drive field [Fig.~\ref{fig:5}(a)]. The drive antenna is placed just beside the YIG sphere [Figs.~\ref{fig:5}(b) and 5(c)] and it is connected to a superconducting microwave line that goes into the cavity via port 3. This makes it efficient to pump the YIG sphere with a drive field.

\section{HAMILTONIAN OF THE COUPLED HYBRID SYSTEM}\label{AppB}

The hybrid system shown in Fig.~\ref{fig:5} consists of a small YIG sphere coupled to a 3D rectangular cavity and driven by a microwave field. Its Hamiltonian can be written as (setting $\hbar=1$)
\begin{equation}\label{sup1}
H=H_{\rm{c}}+H_{\rm{m}}+H_{\rm{int}}+H_{\rm{d}}.
\end{equation}
Here $H_{\rm{c}}=\omega_{\rm{c}}a^{\dag}a$ is the Hamiltonian of the cavity mode $\rm{TE}_{\rm{102}}$ used in our experiment, with  $\omega_{\rm{c}}$ and $a^{\dag}$($a$) being the frequency and creation (annihilation) operator of the cavity mode, respectively. When Zeeman energy, demagnetization energy and magnetocrystalline anisotropy energy are included, the Hamiltonian of the YIG sphere, which has a volume $V_{\rm{m}}$, can be written as~\cite{Blundell}
\begin{equation}\label{sup2}
\begin{split}
H_{\rm{m}}=&-\int_{V_{\rm{m}}}\mathbf{M}\cdot\mathbf{B}_{0}d\tau\\
           &-\frac{\mu_{0}}{2}\int_{V_{\rm{m}}}\mathbf{M}\cdot(\mathbf{H}_{\rm{de}}+\mathbf{H}_{\rm{an}})d\tau
\end{split}
\end{equation}
where $\mu_0$ is the magnetic permeability of free space, $\mathbf{B}_{0}=B_{0}\mathbf{e}_{z}$ is the static magnetic field applied in the $z$ direction which is aligned along the crystalline axis $[100]$ of the YIG sphere in our experiment, $\mathbf{M}$ is the magnetization of the YIG sphere, $\mathbf{H}_{\rm{de}}$ is the demagnetizing field induced by the static magnetic field, and $\mathbf{H_{\rm{an}}}$ is the anisotropic field caused by the magnetocrystalline anisotropy in YIG. For a uniformly magnetized YIG sphere, the induced demagnetizing field is~\cite{Blundell} $\mathbf{H}_{\rm{de}}=-\mathbf{M}/3$, and the anisotropic field is~\cite{Macdonald} $\mathbf{H}_{\rm{an}}=-(2K_{\rm{an}}/M^{2})M_{\rm{z}}$, where only the dominant first-order anisotropy constant $K_{\rm{an}}$ is taken into account and $M$ is the saturation magnetization. Then, the Hamiltonian in Eq.~(\ref{sup2}) becomes
\begin{equation}\label{sup3}
H_{\rm{m}}=-B_{0}M_{z}V_{\rm{m}}+\frac{\mu_{0}}{6}M^{2}V_{\rm{m}}
+\frac{\mu_{0}K_{\rm{an}}}{M^{2}}M_{\rm{z}}^{2}V_{\rm{m}}.
\end{equation}
The YIG sphere can act as a macrospin $\mathbf{S}=\mathbf{M}V_{\rm{m}}/\gamma\equiv(S_{x},S_{y},S_{z})$, where $\gamma=g\mu_{B}/\hbar$ is the gyromagnetic ration~\cite{Soykal}, with $g$ being the g-factor and $\mu_{B}$ the Bohr magneton. With the macrospin operator introduced, the Hamiltonian $H_{\rm{m}}$ reads
\begin{equation}\label{sup4}
H_{\rm{m}}=-\gamma B_{0}S_{z}+\frac{\mu_{0}K_{\rm{an}}\gamma^{2}}{M^{2}V_{\rm{m}}}S_{z}^{2}.
\end{equation}
where we have neglected the constant term $\mu_{0}M^{2}V_{\rm{m}}/6$. The interaction Hamiltonian between the macrospin and the cavity mode is
\begin{equation}
H_{\rm{int}}=g_{\rm{s}}(S^{+}+S^{-})(a^{\dag}+a)\equiv 2g_{\rm{s}}S_{x}(a^{\dag}+a),
\end{equation}
where $g_{s}$ denotes the coupling strength between the macrospin and the cavity mode, and $S^{\pm}\equiv S_{x}\pm iS_{y}$ are the raising and lowering operators of the macrospin, respectively.
In our experiment, the YIG sphere (i.e., the macrospin) is directly pumped by a drive field with frequency $\omega_{\rm{d}}$. The interaction between the macrospin and the drive field is
\begin{equation}
\begin{split}
H_{\rm{d}}=\Omega_{\rm{s}}(S^{+}+S^{-})(e^{i\omega_{\rm{d}}t}+e^{-i\omega_{\rm{d}}t})
    \equiv 4\Omega_{\rm{s}}S_{x}\cos(\omega_{\rm{d}}t),
\end{split}
\end{equation}
where $\Omega_{\rm{s}}$ characterizes the coupling strength of the drive field with the macrospin.

The macrospin operators are related to the magnon operators via the Holstein-Primakoff transformation~\cite{Holstein}:
\begin{eqnarray}
S^{+}&=&\left(\sqrt{2S-b^{\dag}b}\right)b, \nonumber\\
S^{-}&=& b^{\dag}\left(\sqrt{2S-b^{\dag}b}\right), \\
S_{z}&=& S-b^{\dag}b, \nonumber
\end{eqnarray}
where $S$ is the total spin number of the macrospin operator and $b^{\dag}$($b$) is the creation (annihilation) operator of the magnon with frequency $\omega_{\rm{m}}$. For the low-lying excitations with $\langle b^{\dag}b\rangle/2S\ll1$, one has $S^{+}\approx b\sqrt{2S}$, and $S^{-}\approx b^{\dag}\sqrt{2S}$. Then, the Hamiltonian in Eq.~(\ref{sup1}) becomes
\begin{equation}
\begin{split}
H=&\,\,\omega_{\rm{c}}a^{\dag}a+\omega_{\rm{m}}b^{\dag}b+Kb^{\dag}bb^{\dag}b\\
  &+g_{\rm{m}}(a+a^{\dag})(b+b^{\dag})\\
  &+\Omega_{\rm{d}}(b+b^{\dag})(e^{i\omega_{\rm{d}}t}+e^{-i\omega_{\rm{d}}t}),
\end{split}
\end{equation}
where $\omega_{\rm{m}}=\gamma B_{0}-2\mu_{0}K_{\rm{an}}\gamma^{2}S/(M^{2}V_{\rm{m}})$ is the frequency of the magnon mode, $K=\mu_{0}K_{\rm{an}}\gamma^{2}/(M^{2}V_{\rm{m}})$ is a coefficient characterizing the strength of the nonlinear magnon effect,
$g_{\rm{m}}=\sqrt{2S}g_{\rm{s}}$ denotes the magnon-photon coupling strength, and $\Omega_{\rm{d}}=
\sqrt{2S}\Omega_{\rm{s}}$ denotes the coupling strength of the drive field with the magnon mode.
In the rotating-wave approximation, the Hamiltonian is reduced to
\begin{equation}\label{sup6}
\begin{split}
H=&\,\,\omega_{\rm{c}}a^{\dag}a+\omega_{\rm{m}}b^{\dag}b+Kb^{\dag}bb^{\dag}b+g_{\rm{m}}(a^{\dag}b+ab^{\dag})\\
  &+\Omega_{\rm{d}}(b^{\dag}e^{-i\omega_{\rm{d}}t}+be^{i\omega_{\rm{d}}t}).
\end{split}
\end{equation}
Note that because the YIG sphere contains a very large number of spins, the condition $\langle b^{\dag}b\rangle/2S\ll1$ for the low-lying excitations can be easily satisfied~\cite{You-15}, even when considerable magnons are generated by the drive field.

\section{EFFECTIVE HAMILTONIAN IN THE DISPERSIVE REGIME}\label{AppC}

For convenience of calculations, we first transform the Hamiltonian $H$ in Eq.~(\ref{sup6}) to a rotating reference
frame with respect to the frequency of the drive field by the unitary transformation
\begin{equation}\label{sup7}
R_{1}=\exp(-i\omega_{\rm{d}}a^{\dag}at-i\omega_{\rm{d}}b^{\dag}bt),
\end{equation}
i.e.,
\begin{equation}\label{sup8}
\begin{split}
H'=&\,\,R_{1}^{\dag}HR_{1}-iR_{1}^{\dag}\frac{\partial R_{1}}{\partial t} \\
  =&\,\,\omega_{\rm{c}}a^{\dag}a+\omega_{\rm{m}}b^{\dag}b+Kb^{\dag}bb^{\dag}b+g_{\rm{m}}(a^{\dag}b+ab^{\dag})\\
   &+\Omega_{\rm{d}}(b^{\dag}+b)-(\omega_{\rm{d}}a^{\dag}a+\omega_{\rm{d}}b^{\dag}b) \\
  =&\,\,\delta_{\rm{c}}a^{\dag}a+\delta_{\rm{m}}b^{\dag}b+Kb^{\dag}bb^{\dag}b+g_{\rm{m}}(a^{\dag}b+ab^{\dag})\\
   &+\Omega_{\rm{d}}(b^{\dag}+b),
\end{split}
\end{equation}
with $\delta_{\rm{c}(\rm{m})}\equiv\omega_{\rm{c}(\rm{m})}-\omega_{\rm{d}}$. Here the coupled hybrid system is in the strong coupling regime, i.e., $g_{\rm{m}}\gg\kappa,\gamma_{\rm{m}}$,
where $\kappa$ ($\gamma_{\rm{m}}$) is the decay rate of the cavity (magnon) mode.
The Hamiltonian (\ref{sup8}) can be divided into two parts, $H'=H_{0}+H_{I}$, with the free part
\begin{equation}\label{sup9}
H_{0}=\delta_{\rm{c}}a^{\dag}a+\delta_{\rm{m}}b^{\dag}b+Kb^{\dag}bb^{\dag}b+\Omega_{\rm{d}}(b^{\dag}+b),
\end{equation}
and the interaction part
\begin{equation}\label{sup10}
H_{I}=g_{\rm{m}}(a^{\dag}b+ab^{\dag}).
\end{equation}

Below we use a Fr\"{o}hlich-Nakajima transformation to reduce the Hamiltonian $H'$. It needs to find a unitary transformation $U=\exp(V)$, where $V$ is
an anti-Hermitian operator $V^{\dag}=-V$ and satisfies $[H_{0},V]+H_{I}=0$. Up to the second order, the reduced Hamiltonian is given by
\begin{equation}\label{sup11}
H'_{\rm{eff}}=U^{\dag}H'U\approx H_{0}+\frac{1}{2}[H_{I},V].
\end{equation}
We choose $V=\lambda_{1}(a^{\dag}b-ab^{\dag})+\lambda_{2}(a^{\dag}-a)$. Then,
\begin{widetext}
\begin{equation}\label{sup12}
\begin{split}
[H_{0},V]+H_{I}=&\,\,\Big[\delta_{\rm{c}}a^{\dag}a+\delta_{\rm{m}}b^{\dag}b+Kb^{\dag}bb^{\dag}b+\Omega_{\rm{d}}(b^{\dag}+b),
                  \lambda_{1}(a^{\dag}b-ab^{\dag})+\lambda_{2}(a^{\dag}-a)\Big]+g_{\rm{m}}(a^{\dag}b+ab^{\dag})\\
               =&\,\,\lambda_{1}(\delta_{\rm{c}}-\delta_{\rm{m}})(a^{\dag}b+ab^{\dag})
                -\lambda_{1}K\Big[(2b^{\dag}b+1)a^{\dag}b+ab^{\dag}(2b^{\dag}b+1)\Big]
                -\lambda_{1}\Omega_{\rm{d}}(a^{\dag}+a)\\
                &+\lambda_{2}\delta_{\rm{c}}(a^{\dag}+a)+g_{\rm{m}}(a^{\dag}b+ab^{\dag}).
\end{split}
\end{equation}
\end{widetext}
In our experiment, we use a drive field to directly pump the YIG sphere, so as to generate considerable magnons. In this case, the mean-field approximation can be applied to the term $b^{\dag}b$ in Eq.~(\ref{sup12}).
Because $\langle b^{\dag}b\rangle\gg 1$, Eq.~(\ref{sup12}) can approximately be written as
\begin{equation}
\begin{split}
[H_{0},V]+H_{I}\approx &\,\, \Big[\lambda_{1}(\delta_{\rm{c}}-\delta_{\rm{m}})-2\lambda_{1}K\langle b^{\dag}b\rangle+g_{\rm{m}}\Big](a^{\dag}b+ab^{\dag})\\
                       &+(-\lambda_{1}\Omega_{\rm{d}}+\lambda_{2}\delta_{\rm{c}})(a^{\dag}+a).
\end{split}
\end{equation}
Using the relation $[H_{0},V]+H_{I}=0$, we get
\begin{equation}\label{sup13}
\begin{split}
&\lambda_{1}(\delta_{\rm{c}}-\delta_{\rm{m}})-2\lambda_{1}K\langle b^{\dag}b\rangle+g_{\rm{m}}=0,\\
&-\lambda_{1}\Omega_{\rm{d}}+\lambda_{2}\delta_{\rm{c}}=0,
\end{split}
\end{equation}
which give
\begin{equation}\label{sup14}
\begin{split}
\lambda_{1}&=-\frac{g_{\rm{m}}}{\delta_{\rm{c}}-\delta_{\rm{m}}-2K\langle b^{\dag}b\rangle}\\
           &=-\frac{g_{\rm{m}}}{\Delta-2K\langle b^{\dag}b\rangle},\\
\lambda_{2}&=\frac{\Omega_{\rm{d}}}{\delta_{\rm{c}}}\lambda_{1}
            =-\frac{\Omega_{\rm{d}}}{\delta_{\rm{c}}}\frac{g_{\rm{m}}}{\Delta-2K\langle b^{\dag}b\rangle},
\end{split}
\end{equation}
where $\Delta=\omega_{\rm{c}}-\omega_{\rm{m}}$. Therefore, $V$ has the form
\begin{equation}\label{sup15}
\begin{split}
V=&-\frac{g_{\rm{m}}}{\Delta-2K\langle b^{\dag}b\rangle}(a^{\dag}b-ab^{\dag})\\
  &-\frac{\Omega_{\rm{d}}}{\delta_{\rm{c}}}\frac{g_{\rm{m}}}{\Delta-2K\langle b^{\dag}b\rangle}(a^{\dag}-a).
\end{split}
\end{equation}
Also, we apply the mean-field approximation to the Kerr term in Eq.~(\ref{sup11}). Then, the Hamiltonian
(\ref{sup11}) becomes
\begin{eqnarray}\label{sup16}
H'_{\rm{eff}} &\approx &H_{0}+\frac{1}{2}[H_{I},V] \nonumber\\
             &\approx &\bigg[\delta_{\rm{c}}+\frac{g_{\rm{m}}^{2}}{\Delta}+\frac{2g_{\rm{m}}^{2}}{\Delta^{2}}K\langle b^{\dag}b\rangle\bigg]a^{\dag}a\nonumber\\
              & &   +\bigg[\delta_{\rm{m}}-\frac{g_{\rm{m}}^{2}}{\Delta}
                   +\left(1-\frac{2g_{\rm{m}}^{2}}{\Delta^{2}}\right)K\langle b^{\dag}b\rangle\bigg]b^{\dag}b\nonumber\\
              & &+\Omega'_{\rm{d}}(b^{\dag}+b),
\end{eqnarray}
with
\begin{equation}
\Omega'_{\rm{d}}=\bigg[1-\frac{1}{2\delta_{\rm{c}}}\bigg(\frac{g_{\rm{m}}^{2}}{\Delta}
+\frac{2g_{\rm{m}}^{2}}{\Delta^{2}}K\langle b^{\dag}b\rangle\bigg)\bigg]\Omega_{\rm{d}}.
\end{equation}

Finally, we further rotate the reduced Hamiltonian $H'_{\rm{eff}}$ using the unitary transformation
\begin{equation}\label{sup17}
R_{2}\equiv R_{1}^{\dag}=\exp(i\omega_{\rm{d}}a^{\dag}at+i\omega_{\rm{d}}b^{\dag}bt),
\end{equation}
which is the inverse transformation of $R_{1}$ in Eq.~(\ref{sup7}). The derived Hamiltonian is given by
\begin{eqnarray}\label{sup18}
H_{\rm{eff}}&= &R_{2}^{\dag}H'_{\rm{eff}}R_{2}-iR_{2}^{\dag}\frac{\partial R_{2}}{\partial t} \nonumber\\
       &=&\bigg[\omega_{\rm{c}}+\frac{g_{\rm{m}}^{2}}{\Delta}+\frac{2g_{\rm{m}}^{2}}{\Delta^{2}}K\langle b^{\dag}b\rangle\bigg]a^{\dag}a\nonumber\\
       &  &       +\bigg[\omega_{\rm{m}}-\frac{g_{\rm{m}}^{2}}{\Delta}
                 +\left(1-\frac{2g_{\rm{m}}^{2}}{\Delta^{2}}\right)K\langle b^{\dag}b\rangle\bigg]b^{\dag}b\nonumber\\
       &  &         +\Omega'_{d}(b^{\dag}e^{-i\omega_{d}t}+be^{i\omega_{\rm{d}}t}).
\end{eqnarray}
This is the effective Hamiltonian of the coupled hybrid system obtained in the dispersive regime [i.e., Eq.~(\ref{s})]. In our experiment, the drive field is tuned to be in resonance with the magnon mode,
\begin{equation}
\omega_{\rm{d}}=\omega_{\rm{m}}-\frac{g_{\rm{m}}^{2}}{\Delta}
                 +\left(1-\frac{2g_{\rm{m}}^{2}}{\Delta^{2}}\right)K\langle b^{\dag}b\rangle.
\label{resonance}
\end{equation}

\section{RELATION BETWEEN THE MAGNON FREQUENCY SHIFT AND THE DRIVE POWER}\label{AppD}

With Hamiltonian (\ref{sup8}), we can obtain the quantum Langevin equations for the coupled hybrid system,
\begin{equation}\label{sup19}
\begin{split}
\frac{da}{dt}=&-i\delta_{\rm{c}} a -ig_{\rm{m}}b-\frac{\kappa}{2}a,\\
\frac{db}{dt}=&-i\delta_{\rm{m}} b -i(2Kb^{\dag}b+K)b\\
              &-ig_{\rm{m}}a-i\Omega_{\rm{\rm{d}}}-\frac{\gamma_{\rm{m}}}{2}b.
\end{split}
\end{equation}
Here we write the operator $a$ ($b$) as a sum of the steady-state value and the fluctuation, i.e., $a=A +\delta a$ and $b=B+\delta b$. It follows from Eq.~(\ref{sup19}) that $A$ and $B$ satisfy
\begin{equation}\label{sup20}
\begin{split}
\frac{dA}{dt}=&-i\delta_{\rm{c}} A -ig_{\rm{m}}B-\frac{\kappa}{2}A, \\
\frac{dB}{dt}=&-i\delta_{\rm{m}} B -i(2K|B|^{2}+K)B\\
              &-ig_{\rm{m}}A-i\Omega_{\rm{d}}-\frac{\gamma_{\rm{m}}}{2}B.
\end{split}
\end{equation}
From Eq.~(\ref{resonance}), we have $\delta_{\rm{m}}\equiv\omega_{\rm{m}}-\omega_{\rm{d}}\approx g_{\rm{m}}^{2}/\Delta-K|B|^{2}$, because $\Delta\gg g_{\rm{m}}$ in the dispersive regime.
Also, $\delta_{\rm{c}}\equiv\omega_{\rm{c}}-\omega_{\rm{d}}=\Delta+\delta_{\rm{m}}\approx \Delta$.
At the steady states for both $A$ and $B$, $dA/dt=0$ and $dB/dt=0$. Then, it follows from Eq.~(\ref{sup20}) that
\begin{equation}\label{sup22}
\begin{split}
&-i\Delta A -ig_{\rm{m}}B-\frac{\kappa}{2}A=0,\\
&-i\bigg(K|B|^{2}+\frac{g_{\rm{m}}^{2}}{\Delta}\bigg)B-ig_{\rm{m}}A-i\Omega_{\rm{d}}-\frac{\gamma_{\rm{m}}}{2}B=0.
\end{split}
\end{equation}
Eliminating $A$ in Eq.~(\ref{sup22}), we get
\begin{equation}\label{sup23}
\bigg[K|B|^{2}+\frac{g_{\rm{m}}^{2}}{\Delta}-\frac{g_{\rm{m}}^{2}}{\Delta-i(\kappa/2)}
-i\frac{\gamma_{\rm{m}}}{2}\bigg]B+\Omega_{\rm{d}}=0.
\end{equation}
Because $\Delta\gg\kappa$ and $\gamma_{\rm{m}}\gg\kappa$, Eq.~(\ref{sup23}) is reduced to
\begin{equation}\label{sup24}
\bigg(K|B|^{2}-i\frac{\gamma_{\rm{m}}}{2}\bigg)B+\Omega_{\rm{d}}=0.
\end{equation}
Using Eq.~(\ref{sup24}) and its complex conjugate expression, we obtain
\begin{equation}\label{sup25}
\bigg[\Big(K|B|^{2}\Big)^{2}+\bigg(\frac{\gamma_{\rm{m}}}{2}\bigg)^{2}\bigg]|B|^{2}-\Omega_{\rm{d}}^{2}=0.
\end{equation}
In our experiment, the measured frequency shift of THE magnons is
\begin{equation}\label{sup26}
\Delta_{\rm{m}}=\bigg(1-\frac{2g_{\rm{m}}^{2}}{\Delta^{2}}\bigg)K\langle b^{\dag}b\rangle \approx  K\langle b^{\dag}b\rangle.
\end{equation}
Note that $\langle b^{\dag}b\rangle=|B|^2$ for small fluctuation $\delta b$, corresponding to the case with considerable magnons generated in the YIG sphere.
With $\Delta_{\rm{m}}\approx K|B|^{2}$ and $K\Omega_{\rm{d}}^{2}=cP$, where $P$ is the drive power and $c$ is a constant coefficient, Eq.~(\ref{sup25}) reads
\begin{equation}\label{sup27}
\bigg[\Delta_{\rm{m}}^{2}+\bigg(\frac{\gamma_{\rm{m}}}{2}\bigg)^{2}\bigg]\Delta_{\rm{m}}-cP=0,
\end{equation}
which is the relation between the magnon frequency shift and the drive power [i.e., Eq.~(\ref{eq4})].

\end{document}